\documentclass[a4paper,twoside]{article}

\usepackage{epsfig}
\usepackage{subfigure}
\usepackage{calc}
\usepackage{amssymb}
\usepackage{amstext}
\usepackage{amsmath}
\usepackage{amsthm}
\usepackage{multicol}
\usepackage{pslatex}
\usepackage{apalike}

\graphicspath{{./figs/}}

\usepackage{array}
\newcolumntype{P}[1]{>{\centering\arraybackslash}p{#1}}

\usepackage{url}
\usepackage{enumitem}
\usepackage{dblfloatfix}
\usepackage{SCITEPRESS}     % Please add other packages that you may need BEFORE the SCITEPRESS.sty package.

\makeindex

\begin{document}
\title{Model-based Elaboration of a Requirements and Design Pattern Catalogue for Sustainable Systems}

\author{Christophe Ponsard
\affiliation{CETIC Research Centre, Charleroi, Belgium}
\email{christophe.ponsard@cetic.be}}

\keywords{Requirements Modelling, Systems Engineering, Pattern Catalogue, Template, Sustainability, Circular Economy, Fairness, Case study}

\abstract{Designing sustainable systems involves complex interactions between environmental resources, social impact/adoption, and financial costs/benefits. In a constrained world, achieving a balanced design across those dimensions has become challenging. However a number of strategies have emerged to tackle specific aspects such as preserving resources, improving the circularity in product lifecycles and ensuring global fairness. This paper explores how to capture constitutive elements of those strategies using a modelling approach based on a reference sustainability meta-model and pattern template. After proposing an extension to the meta-modelling to enable the structuring of a pattern catalogue, we highlight how it can be populated on two case studies respectively covering fairness and circularity.
}

\onecolumn \maketitle \normalsize \vfill
%%%%%%%%%%%%%%%%%%%%%%%%%%%%%%%%%%%%%%%%%%%%%%%%%%%%%%%%%%%%%

\section{\uppercase{Introduction}}

Sustainability is a broad and multidimensional concept. To quote the Brundtland report from the Unites Nations, ``sustainable development is a process of change in which the exploitation of resources, the direction of investments, the orientation of technological development; and institutional change are all in harmony and enhance both current and future potential to meet human needs and aspirations'' \cite{UN87}. This wise exploitation of resources for the present and future of all humanity impacts how to design our systems. The ICT system and software are also involved and are both  part of the problem and the solution \cite{Calero15}. Defining software sustainability is equally hard with various proposals expressing its long term operation, function preservation over time, availability or a composite non-functional requirement \cite{Venters14}.

Achieving sustainable design goes beyond the consideration of traditional functional and non-functional requirements because of the need to reason in a transverse way across the environmental, social, financial, personal and temporal dimensions that define sustainability \cite{Penzenstadler14,Kienzle20}. Different aspects are emerging and can be applied more or less widely depending on the system considered, e.g. preserving resources, improving the circularity in product lifecycles and ensuring global fairness. This paper will focus on:
\begin{itemize}[noitemsep]
\item \textit{ensuring fairness} in the resource allocation process. This is reflected by Mahatma Gandhi's quote:``There is enough for everybody's need, but not enough for anybody's greed''. This involves to constrain access to resource without necessary aiming in pure formal equality which is not achievable in practice but for a global equilibrium based on core value like solidarity, e.g.  the healthy help the sick, the adult takes care of young/older, because everybody is likely to go through those various roles. This approach is also being stressed through social or distributive justice \cite{Rawl71,Maiese13}. Defining fairness is equally challenging as sustainability because, depending on cultures, the understandings of fairness differ. Every society has way to assess what is fair and unfair through understandable rules and norms. Essentially, people feel it is fair when those who “deserve” things get them and those who do “not deserve” them do not get them \cite{Wallace16}.
\item \textit{ensuring a Circular Economy (CE)} which is defined as ``an economic system of exchange and production which, at all stages of the life cycle of products (goods and services), aims to increase the efficiency of the use of resources and to reduce the impact on the environment while allowing the well-being of individuals'' \cite{Ademe}. Although it emerged in the 1990's, the development of CE has been gradual, with the emergence of the notion of Cradle-to-Cradle (C2C) around 2002 and the actions of the Ellen MacArthur Foundation \cite{MacArthur09}
\end{itemize}

% bugfix presented earlier for bottom
\begin{figure*}[!b]
\centering
\includegraphics[width=0.9\textwidth]{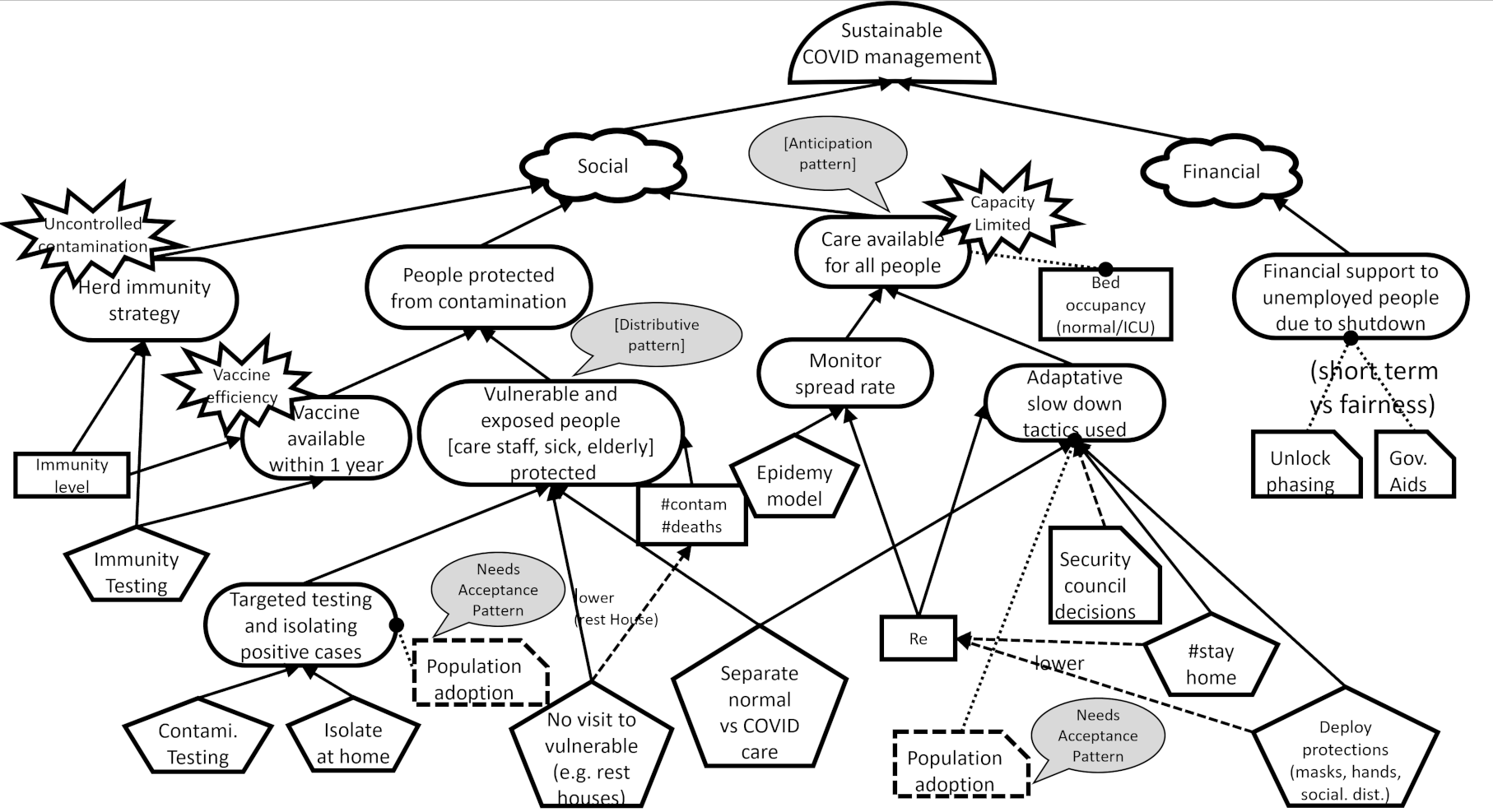}
\caption{COVID supporting case study}
\label{fig:covid-case}
\end{figure*}

In order to reason on sustainability, modelling is clearly advocated \cite{Kienzle20}. Various approaches have been proposed, such as:
\begin{itemize}[noitemsep]
\item generic goal-oriented requirements (GORE) modelling approaches such as i* \cite{Yu97}, KAOS \cite{vlam09} or URN/GRL \cite{URN}, although they may support reasoning, they suffer from their generality and often need specific extensions which make them hard to compose.
\item specific modelling approaches like the Butterfly model \cite{MacArthur09} or the circular value chain \cite{Circulab} used in CE. Although they bring high value, they are hard to compose but are interesting as elicitation tools.
\item specific modelling based on a generic model \cite{Penzenstadler13} or canvas like SuSAF \cite{Duboc20}. However they tend to be lightweight and hard to (de)compose.
\end{itemize}

The purpose of this paper is to build upon the available modelling frameworks to ensure the following requirements:
\begin{enumerate}[noitemsep]
\item capture sustainable systems designs,
\item stay compatible with already adopted frameworks,
\item support decomposition into structural patterns,
\item support their organisation through catalogues similar to OO design patterns \cite{Gamma95}.
\end{enumerate}

% BUGFIX - figures announced here
\begin{figure*}[!t]
\centering
\includegraphics[width=0.9\textwidth]{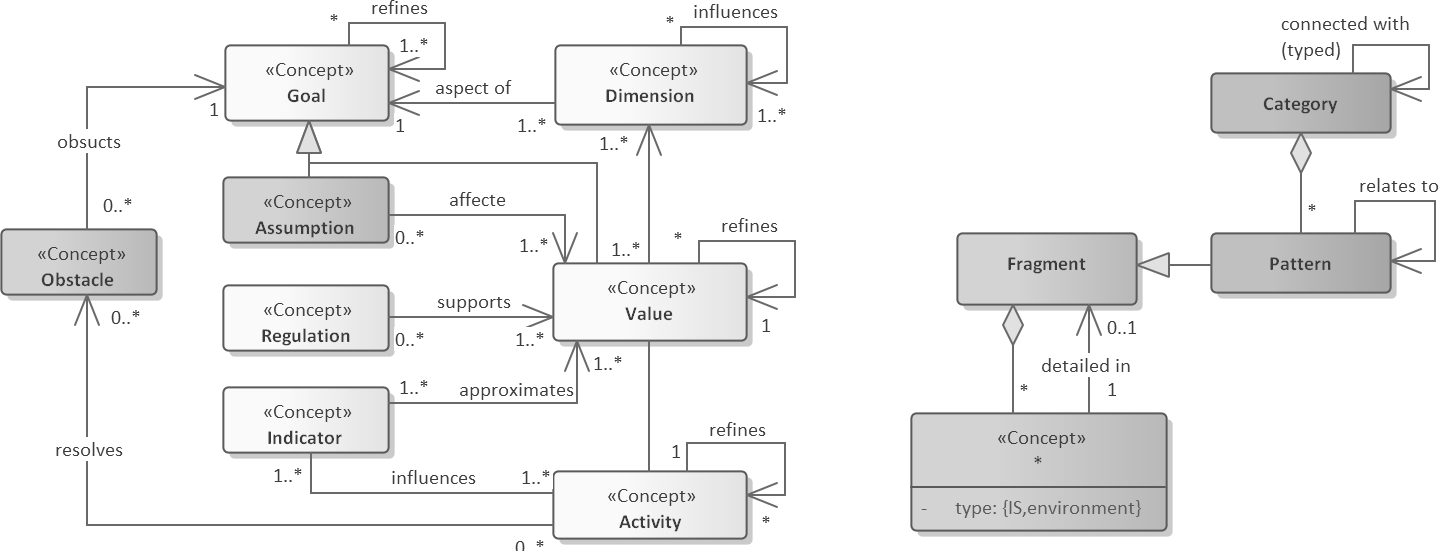}
\caption{Extended meta-model (extension are identified in dark grey)}
\label{fig:mm-extension}
\end{figure*}

\begin{figure*}[!b]
\centering
\includegraphics[width=0.85\textwidth]{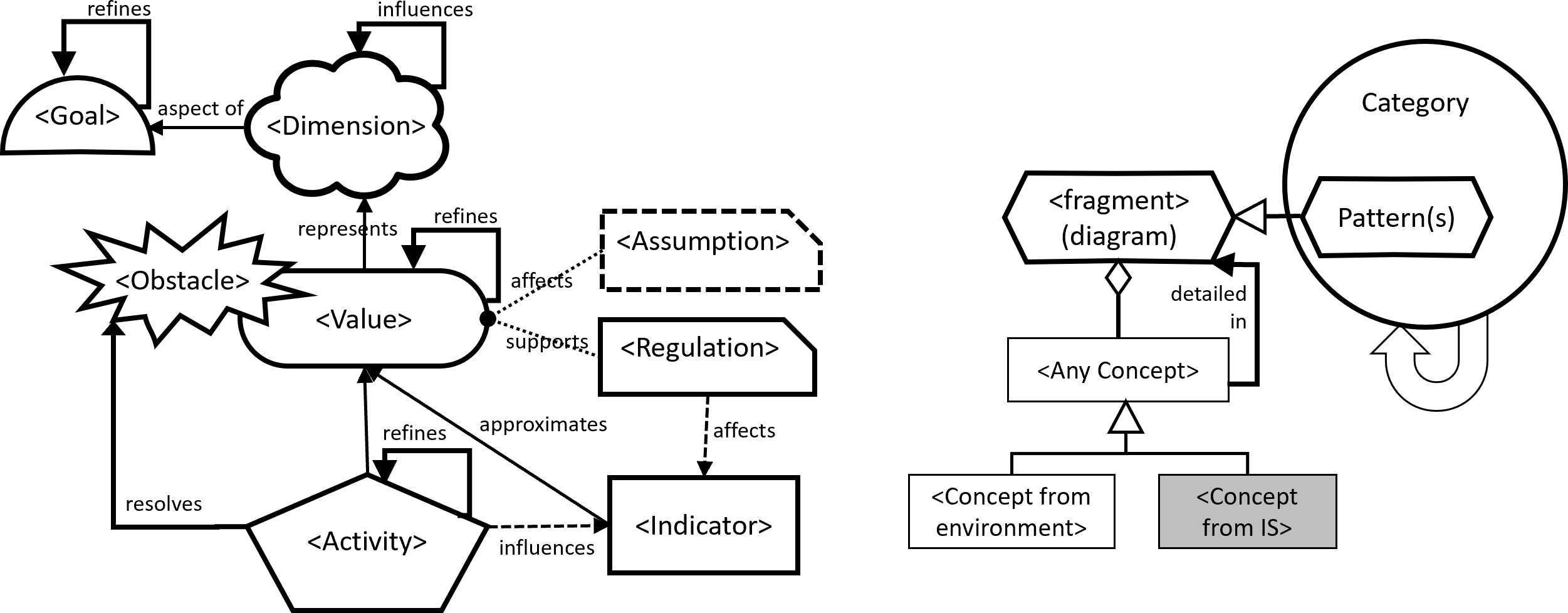}
\caption{Graphical representations for the extended meta-model (note dark grey is used to tag the information system)}
\label{fig:mm-graphical}
\end{figure*}

To achieve this goal, we combined a sustainability reference model \cite{Penzenstadler13} and template \cite{Roher13}. We introduced extensions to support modularity and enable decomposition and isolation of reusable fragments into structured patterns. We also enriched the framework with GORE constructs for more powerful sustainability reasoning especially about robustness of design. The scope of this paper is mostly methodological to show the benefits of a model-based approach, we will not provide extensive details about the produced catalogues nor the possible implementation into supporting tools.

Our paper is structured as follows. Our extension is presented in Section 2 both using a conceptual and graphical meta-model. Its use is validated in Section 3 through two families of strategies resulting in specific catalogues respectively about fairness and circularity. Section 4 then provides a discussion before concluding and identifying further research in Section 5.

\section{\uppercase{Extended Meta-Model for Sustainable Design}}

Our work started by a literature survey of notations used for capturing and reasoning on sustainability, initially with a focus on fairness. We identified about 16 relevant works across different domains such as housing, healthcare, services, logistics and public services \cite{Ponsard21}. Through the use of the early notations and templates proposed by \cite{Penzenstadler14} and \cite{Roher13}, we could cover the first two requirements stated in the introduction about capture and genericity. But in order to cover the next two requirements about structural pattern and catalogue organisation, we had to introduce some extensions presented here both from a conceptual model and the related graphical representation.

Figure \ref{fig:covid-case} illustrates a partial view of a COVID case study which is already quite complex used to introduce the existing conceptual framework and our extensions. It is based on the meta-model and graphical notations respectively depicted in Figures \ref{fig:mm-extension} and \ref{fig:mm-graphical}. The main goal is to achieve sustainable COVID management. It is limited here to social and financial dimensions with a focus on the fairness aspect. On the left of the diagram, some approaches are identified: herd immunity (risky), vaccine (long term) and protective measures which were used in the first stage of the crisis. Note the level of protection is adapted to the vulnerability of target groups which was identified as a general pattern of distributive justice. One extension introduced is the use of bubbles as pattern anchor points. They can help keep track of them in the elicitation phase or after instantiation from our catalogue. However, in this case, relying on a secondary diagram is advised and is also supported by our extension. On the care side, equal access is promoted through the monitoring of the spread of the epidemic and adapting the level of measures. This also relates to a specific anticipation pattern avoiding to reach an unmanageable situation where fair access to care would not be possible any more. Note the use of explosion icons to depict an obstacle to the achievement of a specific activity. This enables reasoning on the robustness of the system. Obstacle resolution tactics can also be captured through specific patterns, e.g. here an anticipation pattern is used to avoid reaching the capacity of the global healthcare infrastructure. Various obstacle resolution strategies have been published by \cite{vlam00}.

\textbf{A key extension was to address the lack of modularity} by introducing a mechanism for breaking down a complex system into smaller parts which are easier to analyse and understand, while preserving a good global picture at the top-level. Our contribution takes the form of a \textit{fragment} concept which can encapsulate a connected subset of the model typically displayed in a specific diagram which limits the visual complexity according to \cite{Moody10}. This fragment can then be linked to a model element in a diagram referencing it. 

\textbf{The use of the fragment extension also enables capturing Design Patterns} i.e. the description of a problem and the essence of its solution to enable the solution to be reused in different settings. In our context, such patterns are collected and organised within a catalogue. Each pattern is described according to the following template based on \cite{Roher13}
\begin{itemize}[noitemsep]
\item \textbf{Summary} - overview stressing the intent, main dimensions and values.
\item \textbf{Category} - name of the primary (and possibly secondary) category of the pattern forming the catalogue typology
\item \textbf{Dimensions} - one or more sustainability dimensions addressed by this pattern. The standard classification covers following dimensions: environmental, economic, social, personal and technical
\item \textbf{Applicability} - context where the pattern is appropriate or not.
\item \textbf{Content} - aspects to be considered for a requirement derived from the pattern, through clear attributes, capabilities, characteristics, or qualities of the system.
\item \textbf{Archetype} - graphical pattern description expressed in a generic way using meta-model concepts and textual explanation.
\item \textbf{Example} - typical instantiation, we will refer to our elicitation case.
\item \textbf{Discussion} - extra details (not developed here due to limited space).
\item \textbf{Related Patterns} - possible interactions or combinations with other patterns.
\end{itemize}

The main additions are the \textit{Category} and \textit{Dimensions} fields. Note that most fields are captured as attributes of the pattern concept except \textit{Category} which is a structuring patterns at the catalogue level, \textit{Archetype} which is a visual representation of its structural content and \textit{Related Patterns} linking together patterns that are typically combined to tackle a specific problem or possible alternatives to a similar problem. 

In order to be able to capture and reason on less idealistic systems, we also introduced some extra notations borrowed from the KAOS framework \cite{vlam09}. First, the notion of \textit{Assumption} enables to make explicit some design choices that could be questioned at a later step, it comes to complete the notion of regulation which is imposed by the environment. Second, the notion of \textit{Obstacle} makes explicit anticipated barriers to the achievement of specific goals or values, by inheritance as depicted in Figure \ref{fig:mm-extension}. Their mitigation can be addressed through specific \textit{Activities} that can materialise strategies such as avoidance/anticipation/repair/degraded mode. Those can be captured and reused through specific patterns. In the COVID case, a violation anticipation is applied to avoid overshooting the hospital capacity. In circular economy, an example is to apply timely maintenance to extend the life of products.

\section{\uppercase{Structuring a Pattern Catalogue}}

This section details the model-based catalogue structuring process and illustrates it on our considered application cases: fairness and circularity. 

\subsection{Gathering and structuring process}

The process of identification and structuring of a pattern catalogue is a long term process that needs to be fed by various sources and actors. At this stage our work is based on the analysis of 16 cases either published in the literature about social housing \cite{Arend20}, sustainable fishing \cite{Doering16}, fair trade \cite{Davenport13}, supply chains \cite{Fearne12}, water management \cite{Syme06}, mobility \cite{Ma18} or produced by our team over the 10 past years in the scope of domain models for various Belgian companies and covering domains such as the design of oncology care, child care, parliamentary systems, taxation, housing, smart building, some of them also published \cite{Ponsard21}.

Each case study was analysed according the following process:
\begin{enumerate}[noitemsep]
\item Identification of values, regulations or assumptions relating to the catalogue scope (fairness or circularity), with attention to specific notions/values such as the balance, distribution or arbitration for fairness or ``looping strategies'' for circular economy.
\item Analysis of existing models based on sustainability or more general notation from requirements or system engineering or process of reconstructing a model from information available at an adequate level of detail.
\item Analysis of business indicators to determine whether they are explicitly or implicitly linked to identified values. 
\item Identification of common/repeating patterns between several case studies, for example, in relation to rules, interactions, adherence process, etc.
\end{enumerate}

The identified patterns are then documented based on their model and according the above template, The category and relation information is used to produce a visual index of the catalogue enabling to identify them within a category and in possible relation with others. Categories may also exhibit more general relations for example based on some temporal/process/logical ordering providing a coherent framework.

\subsection{Fairness Catalogue}

Applying the above process on the available set of cases resulted in the identification of about 12 mature patterns and a few more candidate patterns not included because partially modelled or their reuse could not be assessed to a sufficient extent. As an example, we provide a synthetic version of the previously mentioned violation anticipation pattern using our template. The generalised archetype description is expressed in terms of obstacle control through monitoring capabilities, using some indicators to be identified to track how close to a (fairness) violation we are and activate corrective actions. Its instantiation can be spotted in Figure \ref{fig:covid-case}.

\begin{figure}[h]
    \centering
    \includegraphics[width=\columnwidth]{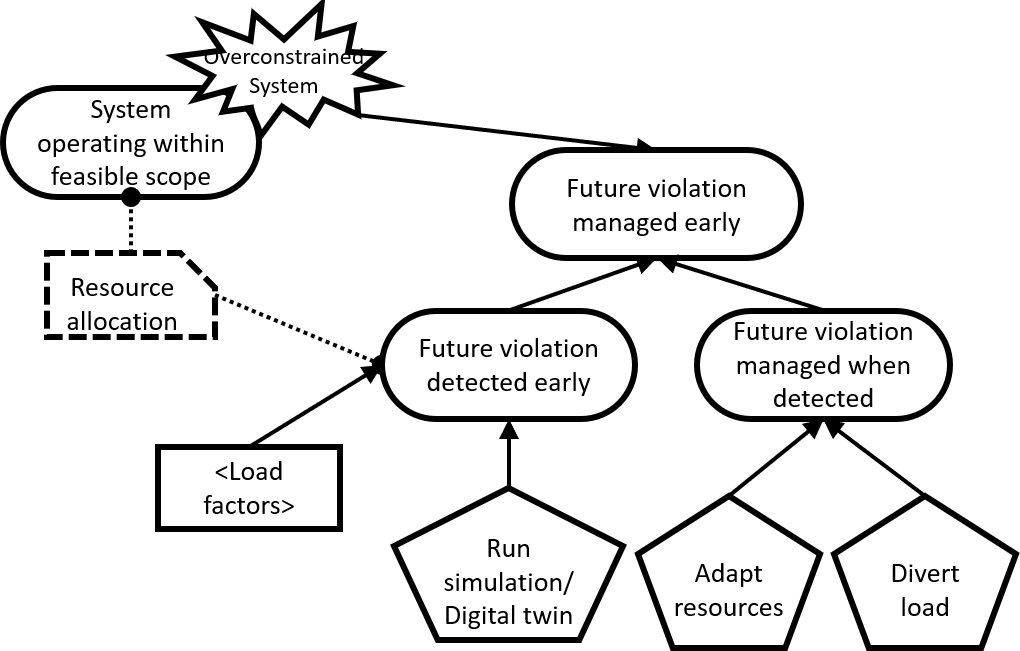}
    \caption{Violation anticipation pattern}
    \label{fig:VAP}
\end{figure}

\begin{itemize}[noitemsep]
\item \textbf{Summary.} This pattern deals with the detection of the violation of a condition (here fairness but applicable more widely). When risk of violation is present, this pattern can anticipate it and thus avoid the occurrence of the violation through an adequate measure.
\item \textbf{Category.} Implementation (primary) and Evolution (secondary).
\item \textbf{Dimensions.} The pattern can be applied to any dimension. It is  mostly used in environmental and social dimensions.
\item \textbf{Applicability.} The considered property (here fairness) must be predicable through some model (related to rules and domain) and collectable data. Enough reaction time to deploy corrective measures is necessary.
\item \textbf{Content.} Predictive model (oracle, simulator, digital twin,...) and data to feed it.
\item \textbf{Archetype}. Figure \ref{fig:VAP} shows the decomposition of the violation pattern intro two steps: detection and management. Detecting relies on the simulation tasks and load factor data. Management can either try to augment resources or to divert the load to make sure it will restore ability to cover all needs and thus not result in unfair situations.
\item \textbf{Example.} Hospital capacity management, human resources.
\item \textbf{Related Pattern.} Rule Acceptance.
\end{itemize}

\vspace{-4mm}

\begin{figure}[!h]
\centering
\includegraphics[width=\columnwidth]{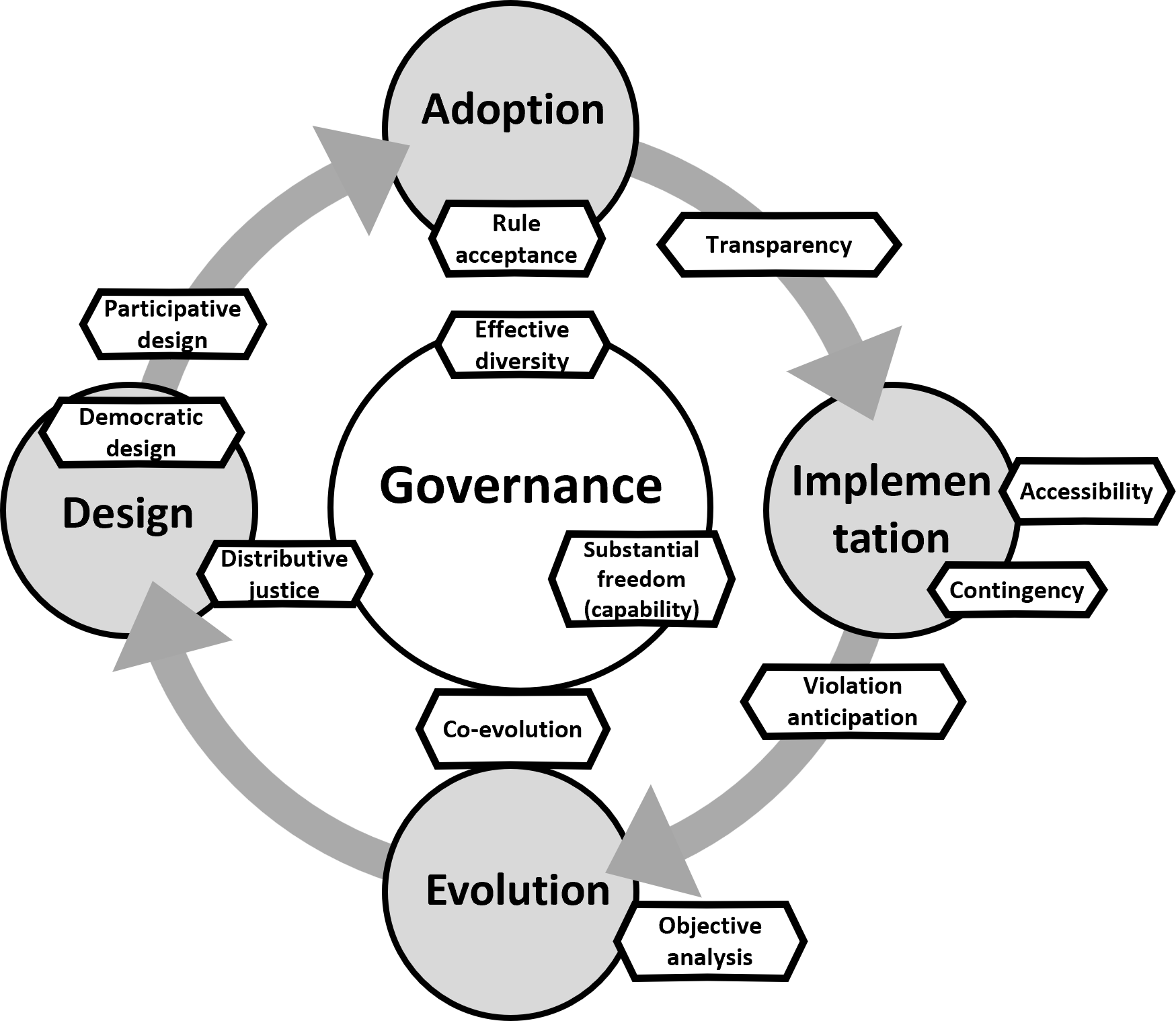}
\caption{Structure of the fairness catalogue}
\label{fig:catalog-fairness}
\end{figure}

The resulting catalogue is depicted in Figure \ref{fig:catalog-fairness}. It is structured across 4 categories chaining to form a virtuous circle that enables a continuous improvement process. This cyclical process is completed by an activity represented at the centre of the cycle about its governance. The identified patterns
are represented on the figure as hexagons placed at a location indicative of their contribution relative to their primary and possibly secondary category, for example co-innovation related both to evolution and governance, transparency to adoption and implementation, while our violation anticipation while mainly at implementation level also enables some form of evolution. At this point, we decided not to include direct links between patterns to avoid overloading the figure, however there is a correlation between those links and the category structure, i.e. related patterns are generally located in the same or a neighbouring category.

\begin{figure*}[!t]
\centering
\includegraphics[width=0.95\textwidth]{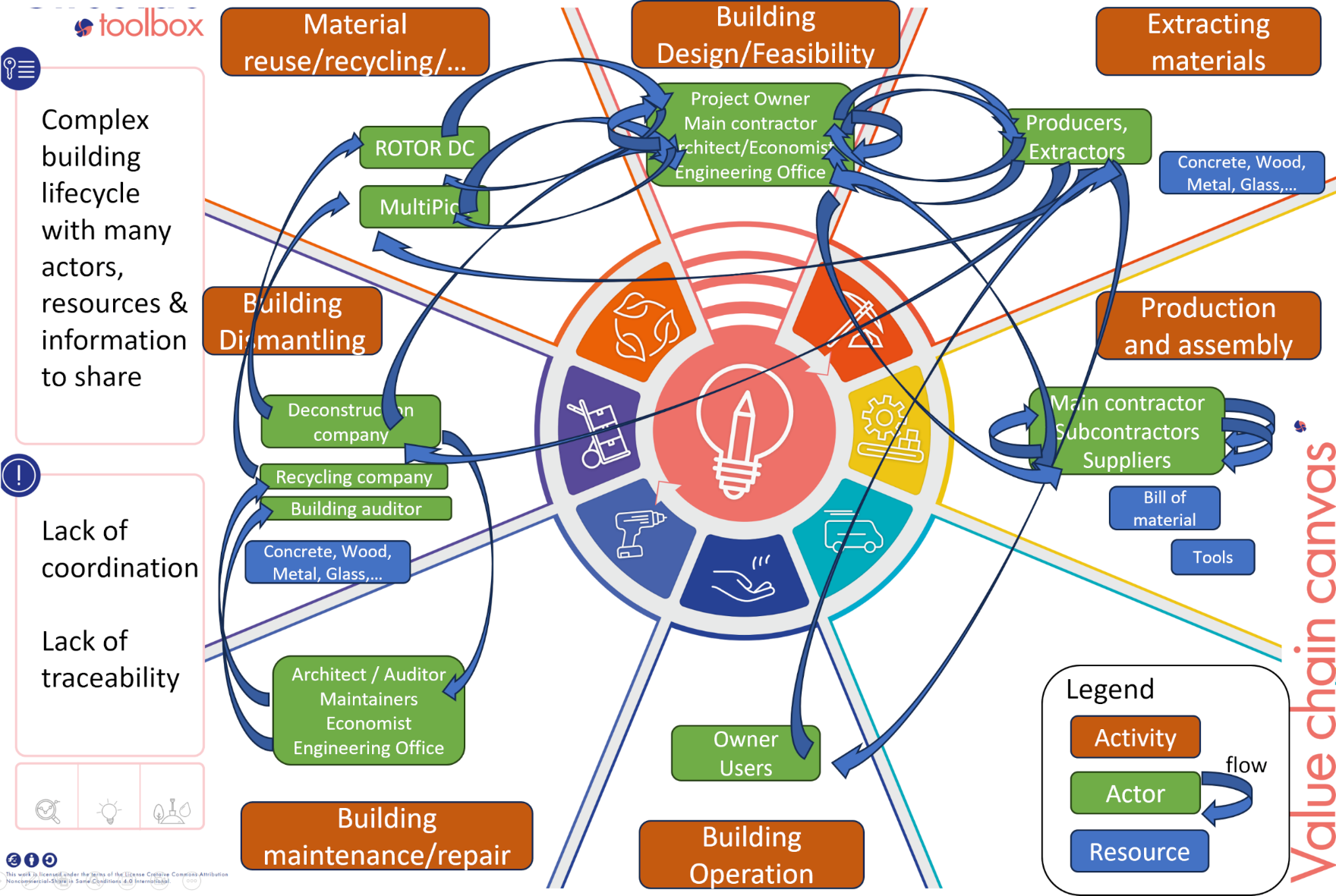}
\caption{Circular value chain canvas}
\label{fig:case-circular}
\end{figure*}

\subsection{Circular Economy Case}

The circular economy case was driven by a specific diagrams used in this domain \cite{Circulab}:
\begin{itemize}[noitemsep]
\item the circular business canvas which enriches the classical business canvas with some extra boxes to capture the synergies around the next usage, various types of resources (natural/technical/energy) and positive/negative impacts.   
\item the circular value chain which is structured around the product lifecycle from design, procurement, construction, usage, dismantling/recycling phases. An example of such a circular value chain is depicted in Figure \ref{fig:case-circular}. It mostly depicts flow of raw to elaborated parts but also of related information (such as material reference and product design models) across different phases. Note that more precise and expressive requirements engineering notations can also be investigated to support this such as i* \cite{Ponsard23b}.
\end{itemize}

\begin{figure}[!h]
\centering
\includegraphics[width=\columnwidth]{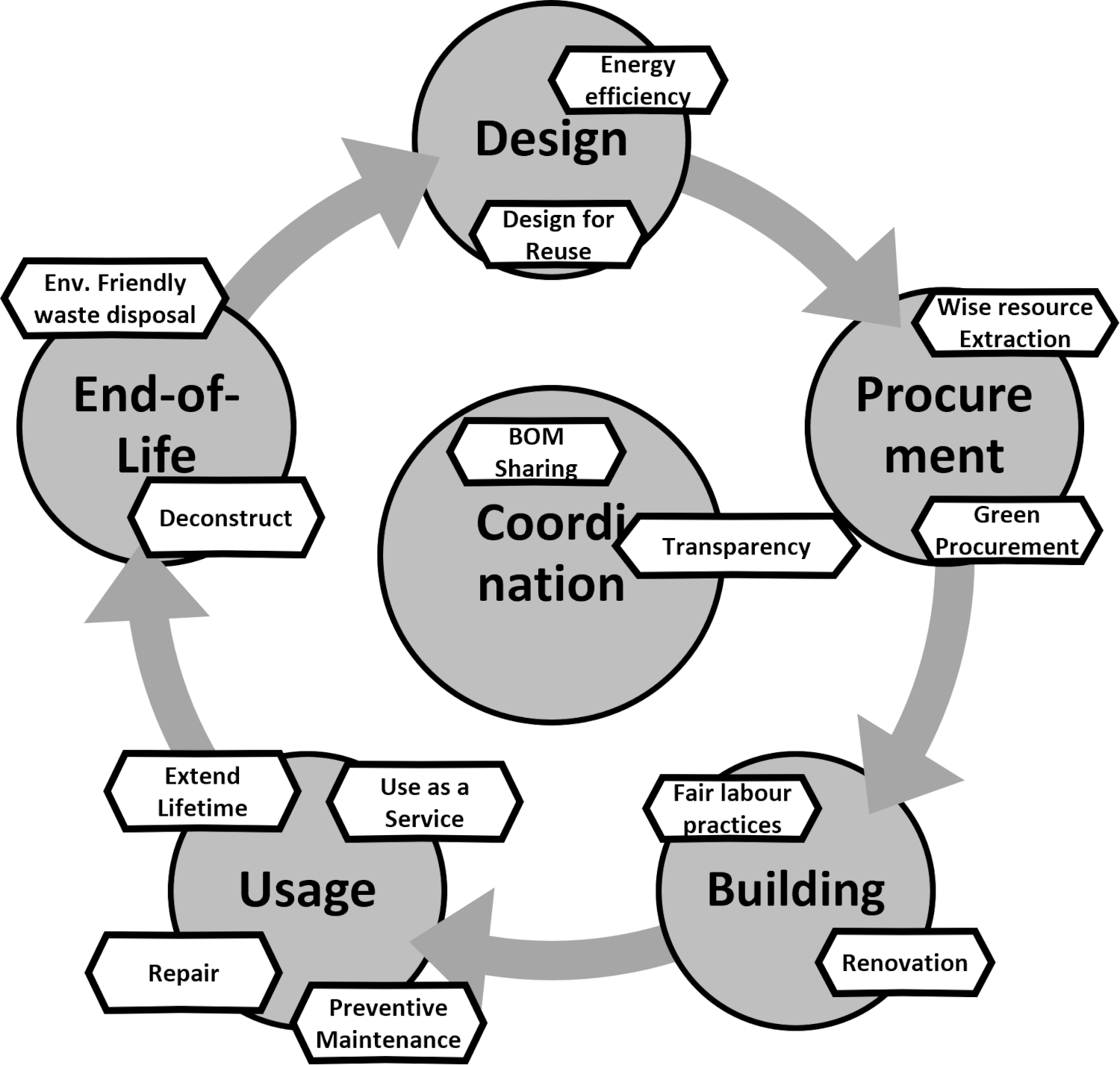}
\caption{Structure of the circular catalogue}
\label{fig:catalog-circular}
\end{figure}

Although both models are interesting, the value chain canvas provided a naturally coherent and easy to understand structuring mechanisms for a pattern catalogue. As in the previous case a central governance category was used to ensure the coordination and information sharing. An analysis similar to the fairness case but to a lesser extent so far has led to the identification of 14 patterns which are depicted in Figure~\ref{fig:catalog-circular}.

Those patterns do not fully capture various R-strategy loops, except for short ones such as repair or life extension. They rather provide building blocks to assemble known strategies or to design new ones. For example, a long loop can be built from design for reuse, then green procurement, renovation built, preventive maintenance and finally easy dismantling. An example of an innovative loop is a more transverse loop relying on design for reuse and transparency to then provide a long term preservation of this information for supporting an efficient dismantling.

\section{\uppercase{Discussion over Related Work}}

Looking back at our work in the light of related work triggered the following feedback. 

About the language itself, it is organised around the key concept of value expressed as a moral or natural good that is perceived as an expression of a specific dimension \cite{Penzenstadler13}. This shares similarities with the concept of goal which captures, at different levels of abstraction, the various objectives the system under consideration should achieve (i.e. prescriptive properties) \cite{vlam09}. Both concepts are system wide, express a notion of measurement of satisfaction and can be refined into more concrete concepts of the same kind. Values may also sound like non-functional requirements (e.g. maintainability, long term use) which is not surprising as one possible definition of sustainability relies on them. The notion of company specific goal is also explicitly used to exemplify the notion of value. One difficulty for an audience familiar with goal-modelling is to avoid using the notation as goal trees. In this respect the proposed patterns are probably not perfect. A check list is to keep in mind the link with the sustainability dimensions, the expression of specific qualities related to sustainability or the focus on the notion of resource. 

So far, our work only considered two different aspects of sustainability (fairness and circularity) in a quite independent way resulting in two different catalogues. Of course, those need to be combined and merged into a wider knowledge repository. This work can be carried out in a bottom-up approach, i.e. progressively extending, generalising, merging available patterns but should be also driven in a top-down manner in order to seek a coherent global picture. Enriching current patterns is not really complex, e.g. for patterns related to fairness, the proposed approach can also cope with other aspects such as equality as considered in \cite{Hinai15}. A natural approach is also to focus on the concept of resource, characterising its nature (renewable, reusable, throwable,...), the transparency of the environmental footprint, the control over its usage (alternatives, avoidance, optimisation, reuse,...). Applying to a wider scope is also interesting to assess the proposed meta-model extension and further enhance it as required while keeping it simple and focused on sustainability reasoning. More specific goal languages may come into play for supporting a broader design analysis of the system to-be. 

In contrast, defining a large scope vision is more difficult and requires more investigation about where to stop and how to provide an usable framework. At the structuring level, a notable inspirational work is the NFR framework which proposes a global organisation of non-functional software requirements \cite{Chung00}. Although not strictly speaking through patterns, it has a systematic organisation and provides many concrete examples. The scope is however limited to software while we consider both the system and identify its IS part enabling more specific analysis and refinement in a later step not covered here. A recent work has also proposed a social and technical sustainability requirements catalogue \cite{Moreira22}. This work is based on a systematic mapping study and relies on the ISO 25000 framework also targeting NFR both for system and software engineering \cite{ISO25K}. The modelling mechanism used is based on feature models and is an interesting alternative to examine although it is quite hierarchical and primarily centred on the ISO standard although a number of interesting additions were made such as the explicit presence of fairness and freedom of risk with mitigation strategies. Those topics are also covered in our own work confirming their relevance.

Finally, in order to be efficient the whole pattern management needs to be supported by a tool covering the capture, classification, retrieval and instantiation phases. While not the focus of this paper, the use of modelling and a MBSE approach strongly ease its implementation of such tools \cite{Schindel2020}. Our current work is in this direction and relies on previous work carried out in smart cards and requirements engineering patterns \cite{Devos12}.

%This partial work certainly requires to improve the documentation of the identified patterns because such kind of library needs to be reviewed and adopted by a whole community. At this stage, most of the work has been devoted to analysis rather than application and it is difficult to estimate the usefulness of the identified patterns. However, we can stress some important factors such as the sharing, assessment of the ROI learning and instantiating vs analysing from scratch, and integrating contributions in a wider body of knowledge. From a research perspective it has already triggered many interesting questions both related to the fairness topic which is the focus of this work but also more generally about the global approach.

% XXX unified catalogue paper XXX

% XXX recent ref survey but istar and no patterns >>>
% A social and technical sustainability requirements catalogue
% Ana Moreira

\section{\uppercase{Conclusion and Perspectives}}

In this paper, we showed how to elaborate catalogues gathering sustainability patterns by gathering a number of available case studies and analysing them using an extended conceptual modelling framework capturing essential requirements, domain properties and also featuring good modularity, reuse and structuring features. Our approach was successfully validated on the construction of two catalogues respectively focusing on fairness and circularity with successful reuse of several patterns on some new or extended cases in the public and health sectors.

This work also triggered the need of further research especially about how to structure larger scale catalogues combining different aspects such as fairness, circularity, resource preservation and how to best combine patterns originating from different aspects. Larger classes of related non-functional properties could also be investigated through the same kind of lens such as robustness which also deals with the long term operation of systems and also requires considering adverse and unexpected conditions.

\noindent \paragraph{Acknowledgements.}

Thanks to RE4SUsy and InforSID communities for feedback and the Belgian construction ecosystem for sharing their case studies through the circular business canvas.

\bibliographystyle{apalike}
\bibliography{modelsward24}

\begin{thebibliography}{}

\bibitem[Mor, 2023]{Moreira22}
 (2023).
\newblock A social and technical sustainability requirements catalogue.
\newblock {\em Data \& Knowledge Engineering}, 143:102107.

\bibitem[{ADEME}, 2014]{Ademe}
{ADEME} (2014).
\newblock Economie circulaire.
\newblock \url{https://www.ademe.fr}.

\bibitem[Calero and Piattini, 2015]{Calero15}
Calero, C. and Piattini, M. (2015).
\newblock {\em Green in Software Engineering}.
\newblock Springer International Publishing.

\bibitem[Chung et~al., 2000]{Chung00}
Chung, L. et~al. (2000).
\newblock {\em Non-Functional Requirements in Software Engineering}.
\newblock Kluwer Academic Publishers.

\bibitem[Circulab, 2012]{Circulab}
Circulab (2012).
\newblock The circulab toolbox.
\newblock \url{https://circulab.com/toolbox-circular-economy}.

\bibitem[Davenport and Low, 2013]{Davenport13}
Davenport, E. and Low, W. (2013).
\newblock From trust to compliance: accountability in the fair trade movement.
\newblock {\em Social Enterprise Journal}.

\bibitem[Devos et~al., 2012]{Devos12}
Devos, N., Ponsard, C., Deprez, J., Bauvin, R., Moriau, B., and Anckaerts, G. (2012).
\newblock Efficient reuse of domain-specific test knowledge: An industrial case in the smart card domain.
\newblock In {\em 34th International Conference on Software Engineering, {ICSE}, Zurich, Switzerland}.

\bibitem[Doering et~al., 2016]{Doering16}
Doering, R., Goti, L., Fricke, L., and Jantzen, K. (2016).
\newblock Equity and itqs: About fair distribution in quota management systems in fisheries.
\newblock {\em Environmental Values}, 25(6):729--749.

\bibitem[Duboc et~al., 2020]{Duboc20}
Duboc, L. et~al. (2020).
\newblock Requirements engineering for sustainability: an awareness framework for designing software systems for a better tomorrow.
\newblock {\em Requirements Engineering}, 25.

\bibitem[Fearne et~al., 2012]{Fearne12}
Fearne, A., Yawson, D., A, B., and J., T. (2012).
\newblock {\em Measuring Fairness in Supply Chain Trading Relationships: Methodology Guide}.

\bibitem[{Fondation Ellen MacArthur}, 2009]{MacArthur09}
{Fondation Ellen MacArthur} (2009).
\newblock Circular economy introduction.
\newblock \url{https://ellenmacarthurfoundation.org/topics/circular-economy-introduction/overview}.

\bibitem[Gamma et~al., 1995]{Gamma95}
Gamma, E., Helm, R., Johnson, R., and Vlissides, J. (1995).
\newblock {\em Design Patterns: Elements of Reusable Object-Oriented Software}.
\newblock Addison-Wesley Longman Publishing Co., Inc., USA.

\bibitem[Hinai and Chitchyan, 2015]{Hinai15}
Hinai, M.~A. and Chitchyan, R. (2015).
\newblock Building social sustainability into software: Case of equality.
\newblock In {\em Fifth {IEEE} Int. Workshop on Requirements Patterns, RePa, Ottawa, ON, Canada, Aug. 25}.

\bibitem[ISO/IEC, 2011]{ISO25K}
ISO/IEC (2011).
\newblock 25010:2011, systems and software engineering — systems and software quality requirements and evaluation (square) — system and software quality models.
\newblock \url{https://iso25000.com/index.php/en/iso-25000-standards/iso-25010}.

\bibitem[{ITU}, 2012]{URN}
{ITU} (2012).
\newblock {Recommendation Z.151 (10/12), User Requirements Notation - Language Def.}
\newblock \url{https://www.itu.int/rec/T-REC-Z.151}.

\bibitem[Jonkman, 2020]{Arend20}
Jonkman, A. (2020).
\newblock {Patterns of distributive justice: social housing and the search for market dynamism in Amsterdam}.
\newblock {\em Housing Studies}.

\bibitem[Kienzle et~al., 2020]{Kienzle20}
Kienzle, J. et~al. (2020).
\newblock Toward model-driven sustainability evaluation.
\newblock {\em Commun. {ACM}}, 63(3):80--91.

\bibitem[Ma et~al., 2018]{Ma18}
Ma, Y., Rong, K., Mangalagiu, D., Thornton, T.~F., and Zhu, D. (2018).
\newblock Co-evolution between urban sustainability and business ecosystem innovation: Evidence from the sharing mobility sector in shanghai.
\newblock {\em Journal of Cleaner Production}, 188.

\bibitem[Maiese, 2013]{Maiese13}
Maiese, M. (2013).
\newblock {Distributive Justice (essay)}.
\newblock \url{https://www.beyondintractability.org/essay/distributive\_justice}.

\bibitem[Moody, 2010]{Moody10}
Moody, D.~L. (2010).
\newblock The "physics" of notations: a scientific approach to designing visual notations in software engineering.
\newblock In {\em ACM/IEEE 32nd International Conference on Software Engineering}, volume~2, pages 485--486.

\bibitem[Penzenstadler et~al., 2014]{Penzenstadler14}
Penzenstadler, B. et~al. (2014).
\newblock Safety, security, now sustainability: The non-functional requirement for the 21st century.
\newblock {\em IEEE Software}, 31(3).

\bibitem[Penzenstadler and Femmer, 2013]{Penzenstadler13}
Penzenstadler, B. and Femmer, H. (2013).
\newblock A generic model for sustainability with process- and product-specific instances.
\newblock In {\em Proc. of the Workshop on Green in/by Software Engineering}, GIBSE ’13, page 3–8.

\bibitem[Ponsard et~al., 2021]{Ponsard21}
Ponsard, C., Nihoul, B., and Touzani, M. (2021).
\newblock Analyse et conception par patrons de l'{\'{e}}quit{\'{e}} de syst{\`{e}}mes d'information durables.
\newblock In {\em Actes du XXXIX{\`{e}}me Congr{\`{e}}s INFORSID, Dijon, France, June 1-4}.

\bibitem[Ponsard and No{\"{e}}l, 2023]{Ponsard23b}
Ponsard, C. and No{\"{e}}l, L. (2023).
\newblock Assessing circular economy ecosystems through i* model analysis.
\newblock In {\em Proc. of the 16th International iStar Workshop}.

\bibitem[Rawls, 1971]{Rawl71}
Rawls, J. (1971).
\newblock {\em A theory of justice}.
\newblock Belknap Press/Harvard University Press.

\bibitem[{Roher} and {Richardson}, 2013]{Roher13}
{Roher}, K. and {Richardson}, D. (2013).
\newblock Sustainability requirement patterns.
\newblock In {\em 3rd Int. Workshop on Requirements Patterns (RePa)}, pages 8--11.

\bibitem[Schindel, 2020]{Schindel2020}
Schindel, W.~D. (2020).
\newblock {\em Pattern-Based Methods and MBSE}.
\newblock Springer International Publishing.

\bibitem[Syme and Nancarrow, 2006]{Syme06}
Syme, G.~J. and Nancarrow, B.~E. (2006).
\newblock Achieving sustainability and fairness in water reform: A western australian case study.
\newblock {\em Water international}, 31(1):23--30.

\bibitem[{United~Nations}, 1987]{UN87}
{United~Nations} (1987).
\newblock {World Commission on Environment and Development: Our Common Future}.
\newblock Oxford Univ. Press.

\bibitem[{van Lamsweerde}, 2009]{vlam09}
{van Lamsweerde}, A. (2009).
\newblock {\em Requirements Engineering - From System Goals to {UML} Models to Software Specifications}.
\newblock Wiley.

\bibitem[van Lamsweerde and Letier, 2000]{vlam00}
van Lamsweerde, A. and Letier, E. (2000).
\newblock Handling obstacles in goal-oriented requirements engineering.
\newblock {\em {IEEE} Trans. Software Eng.}, 26(10):978--1005.

\bibitem[Venters et~al., 2014]{Venters14}
Venters, C. et~al. (2014).
\newblock Software sustainability: The modern tower of babel.
\newblock In {\em {Proc. of the 3rd Int. Workshop on Req. Eng. for Sustainable Systems}}.

\bibitem[Wallace, 2016]{Wallace16}
Wallace, M. (2016).
\newblock {\em From Principle to Practice - A User’s Guide to Do No Harm}.

\bibitem[Yu and Mylopoulos, 1997]{Yu97}
Yu, E. and Mylopoulos, J. (1997).
\newblock Enterprise modelling for business redesign: The i* framework.
\newblock {\em SIGGROUP Bull.}, 18(1).

\end{thebibliography}
%\nocite*

\end{document}